\documentclass[proceedings]{JHEP3}
\usepackage{amssymb}
\usepackage{eurosym}
\usepackage{amsfonts}
\usepackage{amsmath,cite}
\usepackage{epsfig}
\DeclareMathOperator\arccosh{arccosh}
\DeclareMathOperator\arccot{arccot}
\DeclareMathOperator\arcsinh{arcsinh}
\DeclareMathOperator\sech{sech}

\newbox\mybox

\newcommand\fverb{\setbox\mybox=\hbox\bgroup\verb}
\newcommand\fverbdo{\egroup\medskip\noindent\fbox{\unhbox\mybox}\ }
\newcommand\fverbit{\egroup\item[\fbox{\unhbox\mybox}]}
\conference{Infinite series of time-dependent Dyson maps}
\abstract{We propose and explore a scheme that leads to an infinite series of time-dependent Dyson maps which associate different Hermitian Hamiltonians to a uniquely specified time-dependent non-Hermitian Hamiltonian. We identify the underlying symmetries responsible for this feature respected by various Lewis-Riesenfeld invariants. The latter are used to facilitate the explicit construction of the Dyson maps and metric operators. As a concrete example for which the scheme is worked out in detail we present a two-dimensional system of oscillators that are coupled to each other in a non-Hermitian $\mathcal{PT}$-symmetrical fashion.}

\title{Infinite series of time-dependent Dyson maps}
\author{Andreas Fring and Rebecca Tenney \\
Department of Mathematics, City University London,\\
Northampton Square, London EC1V 0HB, UK\\
E-mail: a.fring@city.ac.uk, rebecca.tenney@city.ac.uk}

\input{tcilatex}
\begin{document}

\section{Introduction}
It is well-known for almost thirty years \cite{Urubu} that a given non-Hermitian Hamiltonian $H$ does not uniquely define a quantum mechanical description of a physical system, in the sense that it always leads to a fixed set of well-defined operators associated to physical observables. Even when this Hamiltonian has a set of real eigenvalues and a complete set of eigenstates one needs to construct a well-defined positive-definite metric operator $\rho$ in order to set up a physical Hilbert space. As dictated by the fundamental axioms of quantum mechanics physical observables are then represented by self-adjoint operators acting in this Hilbert space \cite{BB,Alirev,PTbook}. However, the metric $\rho$ is not uniquely determined when only fixing one of the operators associated to observables, e.g. the Hamiltonian $H$, but as was argued in \cite{Urubu}, one needs to specify at least one further observable to render the metric unique. In \cite{Mostsyme} it was shown how for time-independent Hamiltonians the ambiguities in the metric may be related to certain symmetries and for many known models explicit solutions, including some of their ambiguities, have been constructed \cite{JM,PEGAAF,PEGAAF2,Mostsyme,MGH}. Technically one either solves the time-independent quasi-Hermiticity relation $H^{\dagger} \rho = \rho H$ for the metric $\rho$, or uses the fact that the metric factorises into a product of the Dyson map $\eta$ \cite{Dyson} and its conjugate, $\rho = \eta^{\dagger} \eta$ and solves instead the time-independent Dyson equation $\eta H \eta^{-1} = h$ for $\eta$. 

This  pseudo-Hermitian approach has also been extended successfully to non-autonomous systems, that is non-Hermitian explicitly time-dependent Hamiltonian systems \cite{CA,time1,time6,time10,time7,fringmoussa}. In this setting also the metric and the Dyson map become explicitly time-dependent, but the ambiguities persist. In fact, the situation worsens as even the Hamiltonian ceases to be related to a physically observable quantity \cite{fringmoussa}. In the time-independent case it was one of the two natural candidates for operators associated to physical observables that when specified render the metric unique, whereas in the time-dependent case the energy operator has to be defined differently.  

Here our main purpose is to identify the symmetries responsible for associating to a uniquely given non-Hermitian Hamiltonian several equivalent Hermitian Hamiltonians corresponding to different types of physical systems. In particular, we propose and explore a scheme that leads to an infinite series of Dyson maps, hence an infinite series of equivalent Hermitian Hamiltonians, albeit different physics as we shall demonstrate. We employ the Lewis-Riesenfeld method of invariants \cite{Lewis69} to facilitate the explicit construction of the Dyson maps. We explore the working of our proposal for the concrete example of two-dimensional $\mathcal{PT}$-symmetrically coupled harmonic oscillators. 

Our manuscript is organised as follows: In section 2 we explain the general scheme of how a series of infinite Dyson maps can be constructed iteratively from two known solutions to the time-dependent Dyson equation. Section 3 is devoted to setting up the discussion of this scheme for the concrete example of two $\mathcal{PT}$-symmetrically coupled oscillators. We point out a previously overlooked relationship between different types of the underlying auxiliary equations and construct various Lewis-Riesenfeld invariants that are crucial in the understanding of the underlying symmetries as well as being a vital aide in the construction of the time-dependent Dyson maps. In section 4 we construct the infinite series for three different purposefully selected sets of two seed Dyson maps, exploring also the limitations of the scheme. Our conclusions and an outlook into open problems are stated in section 5.

\section{Infinite symmetries and series of Dyson maps from two seeds}

Our starting point is an explicitly non-Hermitian time-dependent Hamiltonian 
$H\neq H^{\dagger }$ satisfying the time-dependent Schr\"{o}dinger equation
(TDSE) $H(x,t)\psi (x,t)=i\hbar \partial _{t}\psi (x,t)$. We further assume
that we have two different time-dependent Dyson maps, $\eta(t) $ and $\tilde{%
\eta}(t)$, satisfying the time-dependent Dyson equations (TDDE) 
\begin{equation}
h=\eta H\eta ^{-1}+i\hbar \partial _{t}\eta \eta ^{-1},\qquad \text{and}%
\qquad \text{ }\tilde{h}=\tilde{\eta}H\tilde{\eta}^{-1}+i\hbar \partial _{t}%
\tilde{\eta}\tilde{\eta}^{-1},  \label{TDDE}
\end{equation}%
involving two different time-dependent Hermitian Hamiltonians $h=h^{\dagger
} $, $\tilde{h}=\tilde{h}^{\dagger }$ that also obey their respective TDSEs 
$h(x,t)\phi (x,t)=i\hbar \partial _{t}\phi (x,t)$ and $\tilde{h}(x,t) \tilde{\phi} (x,t)=i\hbar \partial _{t}\tilde{\phi} (x,t)$. The wavefunctions are related as $\phi= \eta \psi$, $\tilde{\phi} = \tilde{\eta} \psi$ and therefore $\tilde{\phi} = A \phi $, where we employed the first of the operators 
\begin{equation}
A:=\tilde{\eta}\eta ^{-1}  \qquad \text{and}  \qquad \tilde{A}:= \eta ^{-1}\tilde{\eta}. \label{A} 
\end{equation}
The operator $\tilde{A}$ is defined for later purposes. Next we eliminate the Hamiltonian $H$ from the two equations in (\ref{TDDE}), such that
the two Hermitian Hamiltonians are seen to be related as%
\begin{equation}
\tilde{h}=AhA^{-1}+i\hbar \partial _{t}AA^{-1}. \label{gauge}
\end{equation}%
As argued and shown for concrete examples in \cite%
{khantoul2017invariant,maamache2017pseudo,AndTom4,cen2019time,BeckyAndPoint}%
, once the Dyson maps are known one may relate the respective
Lewis-Riesenfeld invariants $I_{\mathcal{H}}$, with $\mathcal{H=}H,h,\tilde{%
	h}$, satisfying \cite{Lewis69}
\begin{equation}
i\hbar \frac{dI_{\mathcal{H}}}{dt}=i\hbar \partial _{t}I_{\mathcal{H}}+[I_{%
\mathcal{H}},\mathcal{H}]=0,   \label{Inv}
\end{equation}%
simply by means of similarity transformations as
\begin{equation}
I_{h}=\eta I_{H}\eta ^{-1},~~\ \ I_{\tilde{h}}=\tilde{\eta}I_{H}\tilde{\eta}%
^{-1}, \quad \Rightarrow  I_{\tilde{h}}=A  I_{h} A^{-1} .\label{inv2}
\end{equation}
Each of the invariants satisfies an eigenvalue equation with time-independent eigenvalues and eigenfunctions that are simply related by a phase factor to the wavefunctions satisfying the respective TDSE. Exploiting the Hermiticity of the invariants $I_{h}$ and $I_{\tilde{h}%
}$, the latter relation in (\ref{inv2}) imply that the operators  
\begin{equation}
S:=A^{\dagger }A \label{S} \qquad \text{and} \qquad \tilde{S}:= A A^{\dagger }
\end{equation}%
are symmetries for the invariants $I_{h}$ and  $I_{\tilde{h}}$, respectively, with
\begin{equation}
\left[ I_{h},S\right] =0 \qquad \text{and} \qquad   \left[ I_{\tilde{h}},\tilde{S}\right] =0 .
\end{equation}%
Thus $S$ and $\tilde{S}$ also satisfy the
Lewis-Riesenfeld equations for the Hermitian $h$-Hamiltonian system and the $\tilde{h}$-Hamiltonian system
\begin{equation}
i\hbar \frac{dS}{dt}=i\hbar \partial _{t}S+\left[ S,h\right] =0, \qquad 
i\hbar \frac{d\tilde{S}}{dt}=i\hbar \partial _{t}\tilde{S}+\left[ \tilde{S},\tilde{h}\right] =0.
\end{equation}
In turn this means that
\begin{equation}
         I'_{h} = I_{h} + S, \qquad \text{and} \qquad I'_{\tilde{h}} = I_{\tilde{h}} + \tilde{S}
\end{equation}
are new invariants for the Hamiltonians $h$ and $\tilde{h}$, respectively.

Another symmetry with an interesting consequence is the possibility of an $\tilde{A}$-symmetry, see (\ref{A}), of the non-Hermitian invariant $I_{H}$, as it implies that the two invariants related to the Hermitian systems are identical
\begin{equation}
   \left[ I_{H}, \tilde{A} \right] =0 	\qquad  \Leftrightarrow   \qquad  I_{h} = I_{\tilde{h}}  ,
\end{equation}
and in turn, the equality of two invariants associated to different Hermitian Hamiltonians implies an $\tilde{A}$-symmetry of the non-Hermitian invariant $I_{H}$. This is easily established by making use of the pseudo-Hermiticity relations for the invariants (\ref{inv2}).

\subsection{Iteration of two Dyson maps}

While certain symmetries of the invariants imply the presence of two inequivalent Dyson maps and vice versa, we will now construct further time-dependent Dyson maps, say $\check{\eta}$ or $\hat{\eta}$, from two given ones, say $\eta $ and $\tilde{\eta}$. We start by constructing a third Dyson map making use of either of the following statements:

\begin{description}
\item[(S1)] 

If and only if the adjoint action of $A$ on the invariant $I_{\tilde{h}}$, $AI_{\tilde{h}}A^{-1}$, is Hermitian then $\check{h}=A\tilde{h}%
A^{-1}+i\hbar \partial _{t}AA^{-1}$ is a new Hamiltonian that is related to
the non-Hermitian Hamiltonian $H$ by the time-dependent Dyson equation $%
\check{h}=\check{\eta}H\check{\eta}^{-1}+i\hbar \partial _{t}\check{\eta}%
\check{\eta}^{-1}$ with $\check{\eta}:=\tilde{\eta}\eta ^{-1}\tilde{\eta}$.

\item[(S2)] 

If and only if the inverse adjoint action of $A$ on the invariant $I_{h}$, $A^{-1}I_{h}A$ is Hermitian then $\hat{h}=A^{-1}hA-i\hbar
A^{-1}\partial _{t}A$ is a new Hamiltonian that is related to the
non-Hermitian Hamiltonian $H$ by the time-dependent Dyson equation $\hat{h}=%
\hat{\eta}H\hat{\eta}^{-1}+i\hbar \partial _{t}\hat{\eta}\hat{\eta}^{-1}$
with $\hat{\eta}:=\eta \tilde{\eta}^{-1}\eta $.
\end{description}

At first we prove (S1) in reverse: Assuming that $\check{\eta}:=\tilde{\eta}%
\eta ^{-1}\tilde{\eta}$ is a new time-dependent Dyson map that maps the non-Hermitian
Hamiltonian $H$ to a Hermitian one, the TDDE $\check{h}=\check{\eta%
}H\check{\eta}^{-1}+i\hbar \partial _{t}\check{\eta}\check{\eta}^{-1}$ holds
by definition. Replacing now $H$ in this equation by means of the first
equation in (\ref{TDDE}) and using the definition (\ref{A}) for $A$, equation $%
\check{h}=A\tilde{h}A^{-1}+i\hbar \partial _{t}AA^{-1}$ follows directly. In
turn this implies that the adjoint action of $A$ on $I_{\tilde{h}}$ yields
the Lewis-Riesenfeld invariant $I_{\check{h}}$. Since $I_{\check{h}}$ is Hermitian, so is $AI_{\tilde{h}}A^{-1}$. The direct statement is
shown by checking whether $AI_{\tilde{h}}A^{-1}$ is Hermitian and then
reversing the steps in the previous argument. Similarly we may prove (S2).

Thus for practical purposes when given the two time-dependent Dyson maps $\eta $, 
$\tilde{\eta}$ and the invariants $I_{\tilde{h}}$, $I_{h}$ we can simply
check whether $AI_{\tilde{h}}A^{-1}$ and/or $A^{-1}I_{h}A$ are Hermitian and
subsequently deduce the form of the new Dyson map. Alternatively one may of
course also assume the given forms for $\check{\eta}$ and $\hat{\eta}$ with
a subsequent check of whether the right hand sides of the corresponding
Dyson equations are Hermitian, thus defining new Hermitian Hamiltonians.

Having now obtained two new time-dependent Dyson maps, we may include them
into the set of the two starting Dyson maps to construct yet more and more
maps by iteration. We summarize the first step as outlines above, i.e. the
constuction of $\check{\eta}=:\eta _{3}$ and $\hat{\eta}=:\eta _{4}$ from
the seed maps $\eta $ and $\tilde{\eta}$, as    
\begin{equation}
\begin{array}{lll}
\eta ,\tilde{\eta} & 
\begin{array}{l}
\nearrow  \\ 
\searrow 
\end{array}
& 
\begin{array}{l}
\eta _{3}=\tilde{\eta}\eta ^{-1}\tilde{\eta}=A\tilde{\eta} \\ 
\\ 
\eta _{4}=\eta \tilde{\eta}^{-1}\eta =A^{-1}\eta 
\end{array}%
.%
\end{array}%
\end{equation}
Replacing now in the next step the seed maps by new maps obtained in the
previous step we obtain, up to the Hermiticity check, 
\begin{eqnarray}
&&%
\begin{array}{lll}
\eta ,\eta _{3} & 
\begin{array}{l}
\nearrow  \\ 
\searrow 
\end{array}
& 
\begin{array}{l}
\eta _{5}=\tilde{\eta}\eta ^{-1}\tilde{\eta}\eta ^{-1}\tilde{\eta}\eta ^{-1}%
\tilde{\eta}=A^{3}\tilde{\eta} \\ 
\\ 
\eta _{6}=\eta \tilde{\eta}^{-1}\eta \tilde{\eta}^{-1}\eta =A^{-2}\eta 
\end{array}%
,%
\end{array}
\\
&&%
\begin{array}{lll}
\eta ,\eta _{4} & 
\begin{array}{l}
\nearrow  \\ 
\searrow 
\end{array}
& 
\begin{array}{l}
\eta _{6}=\eta \tilde{\eta}^{-1}\eta \eta ^{-1}\eta \tilde{\eta}^{-1}\eta
=\eta \tilde{\eta}^{-1}\eta \tilde{\eta}^{-1}\eta  \\ 
\\ 
\tilde{\eta}=\eta \eta ^{-1}\tilde{\eta}\eta ^{-1}\eta 
\end{array}%
,%
\end{array}
\\
&&%
\begin{array}{lll}
\tilde{\eta},\eta _{3} & 
\begin{array}{l}
\nearrow  \\ 
\searrow 
\end{array}
& 
\begin{array}{l}
\eta =\tilde{\eta}\tilde{\eta}^{-1}\eta \tilde{\eta}^{-1}\tilde{\eta} \\ 
\\ 
\eta _{7}=\tilde{\eta}\eta ^{-1}\tilde{\eta}\tilde{\eta}^{-1}\tilde{\eta}%
\eta ^{-1}\tilde{\eta}=\tilde{\eta}\eta ^{-1}\tilde{\eta}\eta ^{-1}\tilde{%
\eta}=A^{2}\tilde{\eta}%
\end{array}%
,%
\end{array}
\end{eqnarray}
\begin{eqnarray}
&&%
\begin{array}{lll}
\tilde{\eta},\eta _{4} & 
\begin{array}{l}
\nearrow  \\ 
\searrow 
\end{array}
& 
\begin{array}{l}
\eta _{7}=\tilde{\eta}\eta ^{-1}\tilde{\eta}\eta ^{-1}\tilde{\eta} \\ 
\\ 
\eta _{8}=\eta \tilde{\eta}^{-1}\eta \tilde{\eta}^{-1}\eta \tilde{\eta}%
^{-1}\eta =A^{-3}\eta 
\end{array}%
,%
\end{array} \\
&&%
\begin{array}{lll}
\eta _{3},\eta _{4} & 
\begin{array}{l}
\nearrow  \\ 
\searrow 
\end{array}
& 
\begin{array}{l}
\eta _{9}=\eta \tilde{\eta}^{-1}\eta \tilde{\eta}^{-1}\eta \tilde{\eta}%
^{-1}\eta \tilde{\eta}^{-1}\eta =A^{-4}\eta  \\ 
\\ 
\eta _{10}=\tilde{\eta}\eta ^{-1}\tilde{\eta}\eta ^{-1}\tilde{\eta}\eta ^{-1}%
\tilde{\eta}\eta ^{-1}\tilde{\eta}=A^{4}\tilde{\eta}%
\end{array}%
.%
\end{array}%
\end{eqnarray}
Continuing in this manner we obtain a series of Dyson maps of the general form%
\begin{equation}
\eta ^{(n)}:=A^{n}\eta ,\qquad \tilde{\eta}^{(n)}:=A^{n}\tilde{\eta},~~\ \
~\ \ \ \ \text{with }n\in \mathbb{Z}.  \label{nDyson}
\end{equation}%
When combined in the way described above we only obtain new maps of the same
form%
\begin{equation}
\begin{array}{lll}
\tilde{\eta}^{(n)},\tilde{\eta}^{(m)} & 
\begin{array}{l}
\nearrow  \\ 
\searrow 
\end{array}
& 
\begin{array}{l}
\tilde{\eta}^{(2m-n)} \\ 
\\ 
\tilde{\eta}^{(2n-m)}%
\end{array}%
,%
\end{array}%
~~~~~~~~~%
\begin{array}{lll}
\quad \tilde{\eta}^{(n)},\eta ^{(m)} & 
\begin{array}{l}
\nearrow  \\ 
\searrow 
\end{array}
& 
\begin{array}{l}
\eta ^{(2m-n-1)} \\ 
\\ 
\tilde{\eta}^{(2n-m+1)}%
\end{array}%
,%
\end{array}%
\end{equation}%
\begin{equation}
\begin{array}{lll}
\eta ^{(n)},\tilde{\eta}^{(m)} & 
\begin{array}{l}
\nearrow  \\ 
\searrow 
\end{array}
& 
\begin{array}{l}
\tilde{\eta}^{(2m-n+1)} \\ 
\\ 
\eta ^{(2n-m-1)}%
\end{array}%
,%
\end{array}%
~~~~~~~~~%
\begin{array}{lll}
\eta ^{(n)},\eta ^{(m)} & 
\begin{array}{l}
\nearrow  \\ 
\searrow 
\end{array}
& 
\begin{array}{l}
\eta ^{(2m-n)} \\ 
\\ 
\eta ^{(2n-m)} 
\end{array}.
\end{array}%
\end{equation}

As discussed above for the iteration to proceed we need to verify at each
step the Hermiticity of the right hand side of the time-dependent Dyson
equation or the adjointly mapped invariants. Thus we require the relevant $A$-operators involving the new maps
\begin{eqnarray}
&&\tilde{\eta}^{(m)} \left( \tilde{\eta}^{(n)} \right)^{-1} = A^{m-n}, \qquad
\eta^{(m)} \left( \tilde{\eta}^{(n)} \right)^{-1} = A^{m-n-1},  \label{An1}\\
&&\tilde{\eta}^{(m)} \left( \eta^{(n)} \right)^{-1} = A^{m-n+1}, \quad
\eta^{(m)} \left( \eta^{(n)} \right)^{-1} = A^{m-n} .  \label{An2}
\end{eqnarray}
Naturally we may repeat the symmetry arguments
from the previous section using the newly constructed Dyson maps, thus obtaining an infinite set of symmetry operators, provided that the Hermiticity property holds at each of the iterative steps.

\section{Two dimensional $\mathcal{PT}$-symmetrically coupled oscillators}

\subsection{Six seed Dyson maps}

To demonstrate the working of the above scheme we shall be considering two-dimensional time-dependent oscillators with a $\mathcal{PT}$-symmetric non-Hermitian coupling in space and momenta. This system has been studied extensively recently \cite{AndTom4,fringtenney21} and is an ideal testing ground for the construction of new Dyson maps as outlined above since six unique solutions have already been identified, that we can build on. The system we will examine is described by the Hamiltonian 
\begin{equation}\label{eq:H}
	H(t) = \frac{a(t)}{2}\left(p_x^2+p_y^2+x^2+y^2\right)+i\frac{\lambda(t)}{2}\left(xy+p_xp_y\right),
\end{equation}
where the time-dependent coefficient functions are taken to be real, \(a(t), \lambda(t) \in \mathbb{R}\). This Hamiltonian is symmetric with regards to two realisations of partial \(\mathcal{PT}\)-symmetries, i.e. \([\mathcal{PT}_\pm,H]=0\), where the anti-linear maps are given by \(\mathcal{PT}_\pm : x \rightarrow \pm x, y \rightarrow \mp y, p_x \rightarrow \mp p_x, p_y \rightarrow \pm p_y, i\rightarrow -i\). It is convenient to express this Hamiltonian in terms of elements of a Lie algebra
\begin{equation}\label{eq:Ha}
	H(t) = a(t)\left(K_1+K_2\right)+i\lambda(t)K_3,
\end{equation}
where the Lie algebraic generators are given by
\begin{equation}
	K_1 = \frac{1}{2}\left(p_x^2+x^2\right), \quad K_2 = \frac{1}{2}\left(p_y^2+y^2\right), \quad K_3 = \frac{1}{2}\left(xy+p_xp_y\right), \quad K_4 = \frac{1}{2}\left(xp_y-yp_x\right),
\end{equation}
satisfying the commutation relations
\begin{align}
	&[K_1,K_2]=0, ~~~~~~~~ [K_1,K_3]=iK_4, \qquad [K_1,K_4] = -iK_3, \notag\\
	&[K_2,K_3] = -iK_4, ~~ [K_2,K_4] = iK_3, \qquad [K_3,K_4] = i(K_1-K_2)/2. \label{Kalgebra}
\end{align}
The six unique Dyson maps which were constructed using a perturbative approach \cite{fringtenney21} map the non-Hermitian Hamiltonian (\ref{eq:Ha}) by means of (\ref{TDDE}) into a general Hermitian Hamiltonian of the form
\begin{equation}\label{eq:h}
	h(t) = f_+(t)K_1+f_-(t)K_2,
\end{equation}
where the time-dependent coefficient functions $f_{\pm}(t)$ are specific for each of the maps. The generic form for all of the Dyson maps consists of two factors given by
\begin{equation}\label{eq:dysongeneral}
	\eta(t) = \exp[\gamma_1(t)q_1]\exp[\gamma_2(t)q_2], 
\end{equation}
where the  \(q_1\), \(q_2\) are operators taken to be generators of the algebra (\ref{Kalgebra}). 
\begin{table}[h]
	\begin{center}
		\begin{tabular}{|l|l||l|l|l|l|}
			\hline
			$\eta_i$&	$q_{1},q_{2}$  & $\gamma _{1}$ & $\gamma _{2}$ &$f_\pm(t)$ &constraint\\ 
			\hline\hline
			$\eta_1$&	$K_{4},K_{3}$ & *& * & $a$ &*\\ \hline
			
			$\eta_2$&	$K_{3},K_{4}$  & $\func{arccosh}\left( \chi \right) $ & $\func{arcsinh%
			}\left( \frac{k}{\chi }\right) $ & $a\pm\frac{k\lambda}{2\chi^2}$&$\chi>1$\\ \hline
			
			$\eta_3$&	$K_{4},iK_{1}$  & $\func{arcsinh}\left( k_{3}\sqrt{1+x^{2}}%
			\right) $ & $-\arctan (x )$ & $a-\frac{\lambda(\pm1+\sqrt{1+(1+x^2)k_3^2})}{2(1+x^2)k_3}$&*\\ \hline
			
			$\eta_4$&$K_{4},iK_{2}$  & $\func{arcsinh}\left( k_{4}\sqrt{1+x^{2}}%
			\right) $ & $\arctan (x )$ & $a+\frac{\lambda(\mp1+\sqrt{1+(1+x^2)k_4^2})}{2(1+x^2)k_4}$&*\\ \hline
			
			$\eta_5$&	$K_{3},iK_{1}$  & $\func{arcsinh}\left( k_{5}\sqrt{1+x^{2}}%
			\right) $ & $-\arccot (x )$ & $a+\frac{\lambda(\pm 1+\sqrt{1+(1+x^2)k_5^2})}{2(1+x^2)k_5}$&*\\ \hline
			
			$\eta_6$&	$K_{3},iK_{2}$  & $\func{arcsinh}\left( k_{6}\sqrt{1+x^{2}}%
			\right) $ & $\arccot (x )$ & $a-\frac{\lambda(\mp 1+\sqrt{1+(1+x^2)k_6^2})}{2(1+x^2)k_6}$&*\\ \hline
		\end{tabular}%
	\end{center}
	\caption{Inequivalent Dyson maps $\eta_i$ with specific operators $q_1$, $q_2$ in the factorisation (\ref{eq:dysongeneral}), and parametrisations for $\protect\gamma _{1}$, $\protect\gamma _{2}$ in terms of the auxiliary functions $\protect\chi $  or $x_i$ together with the time-dependent functions \(f_\pm(t)\) in $h(t)$.}
\end{table}

The time-dependent functions \(\gamma_1(t)\), \(\gamma_2(t)\) are constrained in each case by two coupled first order differential equations. When suitably parametrising the $\gamma_i$, these equation were solved up to some auxiliary equations that may also be solved for a given set of initial conditions. In table 1 we report for a selected set of six Dyson maps, \(\eta_i\), the corresponding parametrisations for the $\gamma_i$, the two operators in the factorisation (\ref{eq:dysongeneral}) and the resulting  
time-dependent coefficient functions, $f_{\pm}(t)$ in (\ref{eq:h}), occurring in the Hermitian counter parts.

 The auxiliary equations governing the time-dependence were found to be
\begin{equation}\label{eq:aux}
	\text{Aux}_1: \quad \ddot{x}_i-\frac{\dot{\lambda}}{\lambda}\dot{x}_i-\lambda^2x_i=0 \qquad \text{and} \qquad \text{Aux}_2: \quad \ddot{\chi}-\frac{\dot{\lambda}}{\lambda}\dot{\chi}-\lambda^2\chi=k^2\frac{\lambda^2}{\chi^3},
\end{equation}
where \( i = 3,4,5,6\). The second equation in (\ref{eq:aux}) is the ubiquitous Ermakov-Pinney equation \cite{Ermakov,Pinney}.

\subsection{Relation between auxiliary equations}
To carry out the discussion as set out in section 2 we have the somewhat unappealing feature that various seed Dyson maps are governed by different types of auxiliary equations. Here we comment briefly on a feature previously not realised, and show that with a different parametrisation also \(\eta_2\) is in fact constrained by the linear second order equation in (\ref{eq:aux}). To demonstrate that this can be achieved we briefly recall how to solve the TDDE for \(\eta_2\), but with a different parametrisation. 

Assuming for this purpose the Dyson map \(\eta_2\) to be of the form (\ref{eq:dysongeneral}) with \(\gamma_1(t), \gamma_2(t)\) unknown and $q_1= K_3$, $q_2 = K_4$, we substitute \(\eta_2\) into the TDDE and find that \(h_2(t)\) is indeed made to be Hermitian if the following coupled first order differential equations are satisfied
\begin{equation}
	\dot{\gamma}_1 = -\lambda \cosh\gamma_2 \qquad \text{and} \qquad \dot{\gamma}_2 = \lambda \tanh\gamma_1\sinh\gamma_2. \label{gammagamma}
\end{equation}
To solve these equations for $\gamma_1$ and $\gamma_2$ we notice first that we can eliminate $\lambda$ and $dt$ from the above equations to give
\begin{equation}
	\gamma_2'(\gamma_1) =-\tanh \gamma_1\tanh \gamma_2(\gamma_1),
\end{equation}
which we solve to
\begin{equation}
	\gamma_2 = \arcsinh(c \sech \gamma_1),
\end{equation}
with $c$ being an integration constant. Previously, \cite{fringtenney21}, we parametrised $\gamma_1 = \arccosh\chi$ which lead to the Ermakov-Pinney equation Aux$_2$ as auxiliary equation. When instead we define $\gamma_1 = \arccosh \sqrt{1+x_2^2}$ and let $c = - 1/k_2$ we find that the central equation to be satisfied is now also Aux$_1$, similarly as for the other cases. We have now a new way of writing the Dyson map $\eta_2$ so that all of the  maps found are governed by the same central equation, with
\begin{equation}\label{eq:eta2new}
	\eta_2: \quad \gamma_1 = \arccosh\sqrt{1+x_2^2}, \quad \gamma_2 = \arcsinh\left(-\frac{1}{k_2\sqrt{1+x_2^2}}\right), \quad f_\pm = \mp \frac{\lambda}{2k_2(1+x_2^2)}.
\end{equation}
Thus we have found a way to convert the nonlinear dissipative Ermakov-Pinney equation given by Aux$_2$ to the linear second order differential equation Aux$_1$ by the relation
\begin{equation}
	\chi = \sqrt{1+x_i^2} \qquad \text{with}  \quad k = -\frac{1}{k_i}.
\end{equation}
This appears to be somewhat miraculous, but one needs to stress here that this is only possible when employing also the first order equations resulting from (\ref{gammagamma}) for the respective variables, i.e.
\begin{equation}
	\dot{x}_2 = - \frac{\lambda \sqrt{1+k_2^2(1+x_2^2)}}{k_2}, \label{first}
\end{equation}
for the new parametrisation. Notice also that the constraint imposed on $\chi$ as reported in table 1 is automatically satisfied with the new parametrisation.

\subsection{Construction of invariants}
As outlined in section 2, in order to construct new Dyson maps we must first calculate invariants for each of the Hermitian Hamiltonians associated to each of the seed Dyson maps \(\eta_i, i=2,\dots,6\). Using the corresponding expressions for $h_i$ we solve equation (\ref{Inv}) and find the invariant
\begin{equation}\label{eq:inv1}
	I_{h_i}(t) = c_1K_1+c_2K_2+c_3\cos\left[c_4-\int^t f_{+-}^i(s) ds\right] K_3-c_3\sin\left[c_4-\int^t f_{+-}^i(s) ds\right] K_4
\end{equation}
where  $f_{+-}^i:= f_+^i - f_-^i$  is the difference between the time-dependent functions in (\ref{eq:h}) occurring in the Hermitian Hamiltonian, and the \(c_1,c_2,c_3,c_4\) are real constants. Notice that in all cases the difference takes on the same form 
\begin{equation}
	f_{+-}^{2,3,4}(t) = - \frac{\lambda}{k_i(1+x_i^2)}= -f_{+-}^{5,6}(t),
\end{equation}
such that the corresponding Hermitian invariants are identical. 

While the cases $i=2,\dots,6$ have been unified, the case $i=1$ still stands out as in this case  \(f_+ = f_-\), so that the invariant in (\ref{eq:inv1}) is rendered time-independent. We therefore need to construct an additional invariant for \(h_1(t)\). To achieve that we need to enlarge the algebra (\ref{Kalgebra}) by six additional elements. To allow for a more compact notation we temporarily denote $K_1\rightarrow K_+^x$, $K_2\rightarrow K_+^y$, $K_3\rightarrow I_+$ and $K_4\rightarrow J_-$, so that the ten Hermitian generators can be written as
\begin{equation}
  	K_\pm^z=\frac{1}{2}(p_z^2 \pm z^2), \quad K_0^z=\frac{1}{2}\{z,P_z \}, \quad J_\pm =\frac{1}{2}(xp_y \pm y p_x), \quad I_\pm =\frac{1}{2}(xy \pm  p_x p_y),
\end{equation}
where $z=x,y$. As laid out in \cite{AndTom4,fring2020time}, we then obtain a closed algebra with non-vanishing commutation relations
\begin{eqnarray}
&&	[K_0^x,K_\pm^x ]=2i K_\mp^x, \quad [K_0^y,K_\pm^y ]=2i K_\mp^y, \quad [K_+^x,K_-^x ]=2i K_0^x, \quad [K_+^y,K_-^y ]=2i K_0^y, \quad \quad  \quad \\
&&   [K_0^x,J_\pm ]=-iJ_\mp, \quad \,\,  [K_0^y,J_\pm ]=iJ_\mp, \quad \,\,\,\,\,\, [K_0^x,I_\pm ]=-iI_\mp, \quad \,\,\,\, [K_0^y,I_\pm ]=-iI_\mp,   \\
&& [K_\pm^x,J_+ ]=\pm iI_\mp, \quad \,\, [K_\pm^y,J_+ ]=\pm iI_\mp, \quad \,\, [K_\pm^x,J_- ]=\mp iI_\pm, \quad \, \, [K_\pm^y,J_- ]=\pm iI_\pm, \qquad \quad \\
&& [K_\pm^x,I_+ ]=\pm iJ_\mp, \,\, \quad [K_\pm^y,I_+ ]=- iJ_\mp, \quad \,\, [K_\pm^x,I_- ]=\mp iJ_\pm, \quad \,\, [K_\pm^y,I_- ]=- iJ_\pm, \\
&& [J_+,J_-]=\frac{i}{2}(K_0^x-K_0^y), \quad [I_+,I_-]=-\frac{i}{2}(K_0^x+K_0^y), \\
&& [J_+,I_\pm]=\pm \frac{i}{2}(K_\mp^x+K_\mp^y), \quad [J_-,I_\pm]=\pm \frac{i}{2}(K_\mp^x-K_\mp^y).
\end{eqnarray}	
The \(\mathcal{PT}\)-symmetry that leaves this algebra invariant manifests itself as
\begin{equation}
\mathcal{PT}_{\pm}: \quad	K_0^{x,y} \rightarrow -	K_0^{x,y}, \quad K_\pm^{x,y} \rightarrow 	K_\pm^{x,y},
	\quad I_\pm \rightarrow - I_\pm, \quad J_\pm \rightarrow J_\pm, \quad i\rightarrow -i. 
\end{equation}
Assuming now that the invariant is also spanned by these generators we found as another solution to (\ref{Inv}) a universal solution for all six cases
\begin{equation}\label{eq:inv3}
	I_{h_i}(t) = \alpha_+(K_+^x+K_-^x)+\beta_+(K_+^x-K_-^x)+\alpha_-(K_+^y+K_-^y)+\beta_-(K_+^y-K_-^y)+\delta_+K_{0}^x+\delta_-K_{0}^y,
\end{equation}
for $i = 1, \ldots, 6$. The time-dependent functions are constrained by
\begin{equation}
	\alpha\pm(t) = \rho_\pm(t)^2, \qquad \beta_\pm(t) = \frac{1}{\rho_\pm(t)^2}+\frac{\dot{\rho}_\pm(t)^2}{f_\pm(t)^2}, \qquad \delta_\pm(t) = -\frac{2\rho_\pm(t)\dot{\rho}_\pm(t)}{f_\pm(t)}, 
\end{equation}
where the auxiliary functions $\rho_\pm$ satisfy the dissipative Ermakov-Pinney equation
\begin{equation}
	\ddot{\rho}_\pm-\frac{\dot{f}_\pm}{f_\pm} \dot{\rho}_\pm +f_\pm^2\rho_\pm =
	\frac{f_\pm^2}{\rho_\pm^3}.
\end{equation}
We will exploit these ambiguities and use which ever invariant is most useful in a certain context.
Noting that the invariant in (\ref{eq:inv1}) is much simpler than the one in (\ref{eq:inv3}), we shall be using it below for the Hermitian Hamiltonians $h_i$ associated with the Dyson maps $\eta_i$, $i=2,\ldots,6$. In turn we shall use the invariant (\ref{eq:inv3}) only for the Hermitian Hamiltonian $h_1$ associated with \(\eta_1\), for which it simplifies further due to the relation \(f_+ = f_-\) that implies \(\rho_+ = \rho_-\).

We also construct the non-Hermitian invariant for the non-Hermitian Hamiltonian $H$ in (\ref{eq:H}) by directly solving equation (\ref{Inv}). We find 
\begin{equation}
I_H = C_1(t) K_1+C_2(t) K_2 +C_3(t) K_3 + i C_4(t) K_4, \label{IH}
\end{equation}
with 
\begin{eqnarray}
 C_1 &=& \frac{c_1}{2} + c_3 \cosh \left( c_4 - \int^t \lambda(s) ds   \right), \quad  C_2 =\frac{c_1}{2} - c_3 \cosh \left( c_4 - \int^t \lambda(s) ds   \right),  \quad\\
 C_3 &=& c_2,  \qquad \qquad  C_4 =2 c_3 \sinh \left( c_4 - \int^t \lambda(s) ds \right) .
\end{eqnarray}
Using the equations in (\ref{inv2}) we may relate the various invariants up to the stated ambiguities. We have verified that the inverse adjoint actions of $\eta$ and $\tilde{\eta}$ on $I_H$ in (\ref{IH}) are indeed invariants for $I_h$ and $I_{\tilde{h}}$, respectively, albeit different from the invariants in (\ref{eq:inv1}) and (\ref{eq:inv3}) up the aforementioned ambiguities. 

\section{Infinite series of Dyson maps from two seeds}

For our non-Hermitian Hamiltonian $H(t)$ in (\ref{eq:H}) we have now a number of seed Dyson maps $\eta_i$ at hand together with their associated Hermitian Hamiltonians $h_i$ and their respective Lewis-Riesenfeld invariants $I_{h_i}$. Thus we can now carry out the scheme laid out in section 2 and construct an infinite series of Dyson maps from two of these seed maps. We will not present here all thirty possibilities that may result from these six maps, but select various examples that exhibit different types of features including an example for which the mechanism breaks down.

\subsection{Seed maps $\mathbf{\eta = \eta_3}$  and $\mathbf{\tilde{\eta} = \eta_4}$ - unitary operator A}

The central operator to compute first is $A$ as defined in (\ref{A}). We start with a simple example for which some of its factors commute. Taking $\eta = \eta_3$, $\tilde{\eta} = \eta_4$ as specified in table 1 and setting $k_3=k_4=k$, $x_3=x_4=x$ we obtain
\begin{eqnarray}
	A = \eta_4\eta_3^{-1} &=& e^{\arcsinh\left(k\sqrt{1+x^2}\right)K_4}e^{i\arctan(x)\left(K_1+K_2\right)}e^{-\arcsinh\left(k\sqrt{1+x^2}\right)K_4}  \\
	&=& e^{i \arctan(x)\left(K_1+K_2\right)}.
\end{eqnarray}
The last equality results from the fact that  \(\left[K_1+K_2, K_4\right] = 0\). According to the statement (S1) in section 2.1 we need to guarantee next that \(AI_{h_4}A^{-1}\) is Hermitian. For the case at hand this is easily seen to be the case as A is a unitary operator and \(I_{h_4}\) is Hermitian. Thus, according to (S1) a new Dyson map is given by
\begin{align}
	\eta^{(1)} = A\eta_4 &= e^{i \arctan(x)\left(K_1+K_2\right)}e^{\arcsinh(k\sqrt{1+x^2})K_4}e^{i\arctan(x)K_2}\notag \\
	&= e^{\arcsinh(k\sqrt{1+x^2})K_4}e^{i \arctan(x)\left(K_1+2 K_2\right)},
\end{align}
which in turn is not unitary. Next we compute the associated Hermitian Hamiltonian from the TDDE (\ref{TDDE}) simply bu substituting into the right hand side all the known quantities
\begin{equation}
	h^{(1)} = \left[a+\frac{\lambda\left(3\sqrt{1+k^2(1+x^2)}-1\right)}{2k(1+x^2)}\right]K_1+\left[a+\frac{\lambda\left(3\sqrt{1+k^2(1+x^2)}+1\right)}{2k(1+x^2)}\right]K_2.
\end{equation}
Using next the relation (\ref{nDyson}) it is now straightforward to calculate the infinite series of Dyson maps. At each step the Hermiticity of the adjoint action of the higher order $A$ operators, as defined in (\ref{An1}), (\ref{An2}), on the Hermitian invariants is guaranteed by the fact that also any power of $A$ is a unitary operator. We find
\begin{equation}
	\eta^{(n)} = A^n \eta_4=e^{\arcsinh(k\sqrt{1+x^2})K_4}e^{i \arctan(x)\left[K_1+(n+1) K_2\right]},
\end{equation}
with corresponding infinite series of Hermitian Hamiltonians
\begin{equation}\label{eq:ham1}
	h^{(n)} = 	h^{(1)} + \frac{(n-1)\lambda \sqrt{1+k^2(1+x^2)} }{k (1+x^2)} (K_1+K_2).	
\end{equation}
In a similar fashion we use the second relation in (\ref{nDyson}) to obtain the new Dyson maps
\begin{equation}
	\tilde{\eta}^{(n)} = A^n \eta_3= e^{\arcsinh(k\sqrt{1+x^2})K_4}e^{-i \arctan(x)\left[(n+1)K_1+ K_2\right]},
\end{equation}
with corresponding Hermitian Hamiltonians
\begin{equation}\label{eq:ham2}
	\tilde{h}^{(n)} = \left( \frac{\lambda}{2k(1+x^2)} \right) (K_2-K_1)  + \left[ a- \frac{(2n+1)\lambda \sqrt{1+k^2(1+x^2)} }{2k (1+x^2)} \right] (K_1+K_2) 
\end{equation}

Since $A$ is a unitary operator the symmetry operator as defined in (\ref{S}) is simply the unit operator, i.e. $S = S^{\dagger} = \mathbb{I}$. Moreover the unitarity of $A$ also implies that the relation between the two Hermitian Hamiltonians (\ref{gauge}) simply becomes a non-Abelian gauge symmetry between two Hermitian Hamiltonians. In this case the metric operators do not to change in the iteration process
\begin{equation}
  \rho^{(n)}={\eta^{(n)^\dagger}}\eta^{(n)} = \eta_4^{\dagger}\eta_4=\rho_4, \qquad \text{and} \qquad \tilde{\rho}^{(n)}= \tilde{\eta}^{(n)^\dagger} \tilde{\eta}^{(n)} = \eta_3^{\dagger}\eta_3=\rho_3.
\end{equation}

\subsection{Seed maps $\mathbf{\eta = \eta_2}$  and $\mathbf{\tilde{\eta} = \eta_3}$ - nonunitary operator A}

Once again we start with the construction of the operator $A$
\begin{equation}
	A:= \eta_3\eta_2^{-1} = e^{\arcsinh\left(k\sqrt{1+x^2}\right)K_4}e^{-i\arctan\left(x\right)K_1}e^{\arcsinh\left(\frac{1}{k\sqrt{1+x^2}}\right)K_4}e^{-\arccosh\left(\sqrt{1+x^2}\right)K_3} ,
\end{equation}
where we have used $\eta= \eta_2$, $\tilde{\eta}= \eta_3$  as defined in table 1 and set \(k_2 = k_3 = k\), such that \(x_2 = x_3 = x\). According to (S1) we need to determine again whether the quantity \(A I_{h_3}A^{-1}\) is Hermitian in order to proceed. A lengthy computation can be avoided here by noting that the Hermitian invariants $I_{h}$ for $h_2$ and $h_3$ are identical. Thus we have 
\begin{equation}
A I_{h_3} A^{-1} = \eta_3 \eta_2^{-1} I_{h_2} \eta_2 \eta_3^{-1} = \eta_3  I_{H}  \eta_3^{-1}= I_{h_3}
=I_{h_3}^{\dagger} = \left( A I_{h_3} A^{-1} \right)^{\dagger}, \label{ha}
\end{equation}
Consequently (S1) is implying that
\begin{equation}
	\eta^{(1)} = A \eta_3,
\end{equation}
constitutes a new Dyson map. With the help of the TDDE (\ref{TDDE}) we determine the corresponding Hermitian Hamitonian to 
\begin{equation}
	h^{(1)} = \left[a-\frac{\lambda\left(1+2\sqrt{1+k^2(1+x^2)}\right)}{2k(1+x^2)}\right]K_1+\left[a-\frac{\lambda\left(2\sqrt{1+k^2(1+x^2)}  -1 \right)}{2k(1+x^2)}\right]K_2.
\end{equation}
As previously, we use the relation (\ref{nDyson}) to calculate the infinite series of Dyson maps. With 
\begin{equation}
	\eta^{(n)} = A^n\eta_3, \qquad \text{and} \quad \tilde{\eta}^{(n)} = A^{-n}\eta_2,  \label{etaeta}
\end{equation}
we can use relation (\ref{ha}) repeatedly to ensure that at each level the adjoint action of the higher order $A$s on the Hermitian invariants is Hermitian. Using the TDDE (\ref{TDDE}) for the new maps we obtain the Hermitian Hamiltonians 
\begin{equation}
	h^{(n)} = \left( \frac{\lambda}{2k(1+x^2)} \right) (K_2-K_1)  + \left[ a- \frac{(n+1)\lambda \sqrt{1+k^2(1+x^2)} }{2k (1+x^2)} \right] (K_1+K_2) 
\end{equation}
from the first map in (\ref{etaeta}) and

\begin{equation}
	h^{(n)} = 	h^{(n)} = \left( \frac{\lambda}{2k(1+x^2)} \right) (K_2-K_1)  + \left[ a- \frac{n\lambda \sqrt{1+k^2(1+x^2)} }{2k (1+x^2)} \right] (K_1+K_2) 
\end{equation}
from the second.

We may now also compute the symmetry operators for \(I_{h_2}\) and \(I_{h_3}\). The symmetry operator is readily written down as
\begin{eqnarray}
	S := A^{\dagger}A = && e^{-\arccosh\left(\sqrt{1+x^2}\right)K_3} e^{\arcsinh\left(\frac{1}{k\sqrt{1+x^2}}\right)K_4} e^{-i\arctan\left(x\right)K_1} e^{\arcsinh\left(k\sqrt{1+x^2}\right)K_4}  \qquad \\ 
	&& \times  e^{\arcsinh\left(k\sqrt{1+x^2}\right)K_4}e^{-i\arctan\left(x\right)K_1}e^{\arcsinh\left(\frac{1}{k\sqrt{1+x^2}}\right)K_4}e^{-\arccosh\left(\sqrt{1+x^2}\right)K_3}. \notag
\end{eqnarray}
Thus we may now explicitly verify the symmetry relation (\ref{nDyson}), best calculated in the form $SI_{h_2}S^{-1} = I_{h_2}$.
Similarly, the symmetry operator for $I_{h_3}$ should be given by $\tilde{S}$, which is indeed the case as we verified explicitly.

\subsection{Seed maps $\mathbf{\eta = \eta_1}$  and $\mathbf{\tilde{\eta} = \eta_2, \eta_3, \eta_4}$ - breakdown of the iteration }

From the previous two examples one might get the impression that the iteration procedure can always be carried out with any two seed Dyson maps. However, this is not the case when the Hermiticity condition does not hold. To verify this we relied in the previous section on the fact that the invariants for the two Hermitian Hamiltonians resulting from the seed maps were identical. This is not the case when involving $\eta_1$ as a seed map and any of the other five maps, as can be seen from (\ref{eq:inv3}) when comparing the functions $f_\pm$. Thus in this case the Hermiticity condition needs to be verified more explicitly  

Let us now carry out the calculation for the seed Dyson maps chosen to be $\eta = \eta_1$ and  $ \tilde{\eta} = \eta_2$. We start from the expression for $A$ 
\begin{equation}
	A := \eta_2\eta_1^{-1} = e^{\arccosh\left(\sqrt{1+x_2^2}\right)K_3}e^{-\arcsinh\left(\frac{1}{k_2\sqrt{1+x_2^2}}\right)K_4}e^{\int^s\lambda(s)dsK_3}
\end{equation}
where $\eta_2$ is defined as in (\ref{eq:eta2new}) and $\eta_1$ as in table 1. Next we compute to quantity $A I_{h_2}A^{-1}$ where \(I_{h_2}\) is given by (\ref{eq:inv3}). After a lengthy calculation  we find that this quantity is non-Hermitian and given by
\begin{eqnarray}
	A I_{h_2} A^{-1} = && \left[ \frac{1}{2}\Gamma_+^{++} +\frac{\kappa}{2k_2}   \right] K_1 \left[ \frac{1}{2}\Gamma_+^{++} -\frac{\kappa}{2k_2}   \right] K_2 + i\frac{\cosh(g)}{k_2\Delta \sqrt{1+x_2^2}}\Gamma_-^{+-}K_3 \\
	&&+i\frac{\kappa(1+x_2^2)+\cosh (g)\Delta}{k_2x_2\sqrt{1+x_2^2}}\Gamma_-^{+-}K_4  \notag \\
	&& + i \left\{ \frac{(1+k_2^2)(\delta_-+\delta_+)}{k_2^2x_2}\cosh(g) K_5
	-\frac{1}{k_2^2x_2}\left[k_2x_2 \Gamma_-^{-+} -\kappa\left(\delta_-+\delta_+\right)  \Delta \right] \right\}K_5 \notag \\ 
	&& + i \left\{ \frac{(1+k_2^2) \Gamma_+^{--} }{k_2^2x_2}\cosh(g) 
	 +  \frac{1}{k_2^2x_2}\left[k_2x_2(\delta_--\delta_+)+\kappa \Gamma_+^{--}  \Delta \right] \right\} K_6\notag \\ 
	&& -\frac{\left(\delta _-+\delta _+\right)   \Delta-\Gamma_+^{--} \left(\left(k_2^2+1\right) x_2-k_2^2 x_2\right)}{2 k_2^2 x_2 \sqrt{x_2^2+1}}\cosh(g) (K_7+K_8)\notag \\
	&&+ \left[ \kappa\frac{\delta _-+\delta _+ -x_2 \Gamma_+^{--}   \Delta}{2 k_2^2 x_2 \sqrt{x_2^2+1}} 
	+\frac{\left(\delta _+-\delta _-\right) x_2-\Gamma_-^{+-}   \Delta}{2 k_2 \sqrt{x_2^2+1}} \right]K_7
	\notag \\	
	&&+ \left[ \kappa\frac{-\delta _--\delta _+ -x_2 \Gamma_+^{--}   \Delta}{2 k_2^2 x_2 \sqrt{x_2^2+1}}
	+\frac{\Gamma_-^{+-}   \Delta+\left(\delta _+-\delta _-\right) x_2}{2 k_2 \sqrt{x_2^2+1}} \right] K_8\notag \\
	&&+\frac{ \Gamma_+^{--}   \Delta-\left(\delta _-+\delta _+\right) \left(\left(k_2^2+1\right) x_2-k_2^2 x_2\right)}{2k_2^2 x_2 \sqrt{x_2^2+1}}\cosh(g)((K_9+K_{10})\notag \\
	&&+ \left[ \kappa\frac{\Gamma_+^{--}-\left(\delta _-+\delta _+\right) x_2   \Delta}{2 k_2^2 x_2 \sqrt{x_2^2+1}}
	+\frac{\left(\delta _--\delta _+\right)   \Delta+x_2 \Gamma_-^{+-}}{2 k_2 \sqrt{x_2^2+1}} \right] K_9 \notag \\
	&&+ \left[- \kappa \frac{\Gamma_+^{--}+\left(\delta _-+\delta _+\right) x_2   \Delta}{2 k_2^2 x_2 \sqrt{x_2^2+1}} 
	+ \frac{\left(\delta _+-\delta _-\right)   \Delta+x_2 \Gamma_-^{+-}}{2 k_2 \sqrt{x_2^2+1}}\right]K_{10}, \notag
\end{eqnarray}
where we introduced the abbreviations 
\begin{equation}
 \Gamma_{\delta_1}^{\delta_2\delta_3}:= \alpha_- + \delta_1 \alpha_+ + \delta_2 \beta_- + \delta_3 \beta_+ \quad \Delta:=\sqrt{1+k_2^2(1+x_2^2)} \quad	g :=\int^t\lambda(s)ds,
\end{equation}
 with $\delta_i = \pm 1$, $i=1,2,3$. We simplified here our expressions using the identity 
 \begin{equation}
 	\sinh(g) = \frac{\kappa+\cosh(g)\sqrt{1+k_2^2(1+x_2^2)}}{k_2x_2},
 \end{equation}
which is verified using the first order constraint (\ref{first}).
As the invariant is non-Hermitian the iteration process breaks down and by (S1) we deduce that $A\eta_2$ is not a Dyson map. We have also carried out the equivalent calculation for the seed map choices $\eta = \eta_1$ and  $ \tilde{\eta} = \eta_3$ and $\eta_4$, reaching the same conclusion.

\section{Conclusions}
We proposed a scheme that allows to compute new time-dependent Dyson maps from two seed maps in a iterative fashion for a given non-Hermitian time-dependent Hamiltonian. As argued in general in section 2, in principle the iteration process might continue indefinitely, thus leading to an infinite series of time-dependent Dyson maps including their associated Hermitian Hamiltonians. The symmetry operators $S$ of the Lewis-Riesenfeld invariants govern this behaviour. Thus when the symmetry is broken also the iteration procedure breaks down. We carried out the procedure in detail for a two-dimensional system of harmonic oscillators that are coupled to each other in a non-Hermitian, but $\mathcal{PT}$-symmetrical, fashion. We only presented here three examples that exhibit different types of behaviours, but we have verified that similar results are obtained when starting from different sets of seed functions. We have focused in our analysis mainly on the relations between the various Hamiltonians and their corresponding invariants, but having obtained the Dyson maps, and therefore the metric operators, it is straightforward to extend the considerations to the associated wave functions and inner product structures on the physical Hilbert space.  

Naturally there are various open issues and extensions possible. As always one may carry out the analysis for more concrete examples. Here we were mainly interested in demonstrating the working of the scheme and selected a model that has a simple two dimensional harmonic oscillator as Hermitian counterpart. It would be interesting to carry out the analysis for an example that has more involved Hermitian counterparts. It is also possible to extend the scheme to start with three or more inequivalent seed Dyson maps leading to new maps of different types, see \cite{thesisBecky} for details. In addition, one may also try to generalise the seed functions themselves and start from expressions involving more than the two factors as in the Ansatz (\ref{eq:dysongeneral}).  
\medskip

 \textbf{Acknowledgments:} RT is supported by a City, University of London
Research Fellowship.

\bibliographystyle{phreport}
\bibliography{acompat,Ref}

\end{document}